\documentclass[proof]{WileyASNA-v1}

\articletype{Article Type}%

\received{1 October 2020}
\revised{12 October 2020}
\accepted{25 October 2020}

\begin{document}

\title{Magnetic field decay in young radio pulsars }

\author[1]{A.P. Igoshev}
\author[2,3]{S.B. Popov*}

\authormark{A.P. Igoshev, S.B. Popov}

\address[1]{\orgdiv{Department of Applied Mathematics}, \orgname{University of Leeds}, \orgaddress{\state{Leeds LS2 9JT}, \country{UK}}}

\address[2]{\orgdiv{Sternberg Astronomical Institute}, \orgname{Lomonosov Moscow State University}, \orgaddress{\state{Moscow, 119234}, \country{Russia}}}

\address[3]{\orgdiv{Department of Physics}, \orgname{Higher School of Economics}, \orgaddress{\state{Moscow, 101000}, \country{Russia}}}

\corres{*Moscow, 119234, Universitetsky prospekt 13  \email{polar@sai.msu.ru}}

\abstract{ The role of magnetic field decay in normal radio pulsars is still debated. In this paper we present results which demonstrate that an episode of magnetic field decay in hot young neutron stars can explain anomalous values of braking indices recently measured for more than a dozen of sources. It is enough to have few tens of per cent of such hot NSs in the total population to explain observables. Relatively rapid decay operates at ages $\lesssim$~few~$\times100$~kyrs with a characteristic timescale of a similar value. We speculate that this decay can be related to electron scattering off phonons in neutron star crusts. This type of decay saturates as a neutron star cools down. Later on, a much slower decay due to crustal impurities dominates.   Finally, we demonstrate that this result is in agreement with our early studies. }


\maketitle

\section{Introduction}

Magnetic field evolution in normal radio pulsars is a long-standing problem  (e.g., \citealt{1990ApJ...363..597S} and references therein). 
Some studies (e.g., \citealt{2004ApJ...604..775G, 2010MNRAS.401.2675P, 2015MNRAS.454..615G}) came to the conclusion that moderate decay might happen, as they obtained better fits with slightly decreasing magnetic fields with initial values $B_0\sim10^{12}$~--~$10^{13}$~G.
On other hand, population analysis of radio pulsar properties demonstrates that there is no strong field evolution 
(by a factor about few or more) during the whole lifetime of these sources, and data can even be fitted under the assumption of constant field (e.g., \citealt{2006ApJ...643..332F}).  Still, can it be that the field decays in a non-uniform manner, and some episodes of significant field decrease can happen along the life track? 

In \cite{igoshev2014} we used a modified pulsar current technique to probe history of magnetic field evolution in young neutron stars (NSs). We obtained that the field decayed on average by a factor $\sim 2$ in the brief time interval $\sim 10^5$~--~$\sim3 \times 10^5$~yrs.\footnote{Our method could not probe earlier evolution, as for young objects uncertainty in the initial spin period plays a significant role. Also, we could not analyse data for much older sources, as due to evolution their luminosity becomes significantly smaller and thus, there is a deficit of old pulsars in a sensitivity limited data sample.} Such rapid decay is probably terminated at later stages, as if it continues for more than several hundred thousand years, then the number of visible old pulsars becomes too small in contradiction with observations.

The nature of the termination of relatively rapid field evolution was analysed in \cite{IgoshevPopov2015}.
We considered two main possibilities: NS cooling (which results in diminishing of the resistivity related to electron scattering off phonons) and the Hall attractor. 
In \cite{IgoshevPopov2015} we concluded that the first variant is more probable.
In our new study \citep{2020MNRAS.tmp.2868I} we focused on the fact that most of reliably measured braking indices for young radio pulsars in a new sample \citep{listBrakingInd} are positive  and deviate significantly from a classical value of three, typical for both magneto-dipolar braking in vacuum and filled magnetosphere. Positive braking indices indicate that magnetic field might be decaying and this decay also seems to terminate at timescales comparable to 100 kyr.
So, this decay is somewhat similar to the one identified by \cite{igoshev2014, IgoshevPopov2015}. In addition, in the new study we demonstrate that it can be just a fraction of young NSs which experiences significant field decay due to electron scattering off phonons. 

In this paper we discuss how the results from \cite{2020MNRAS.tmp.2868I} are compatible with our early findings and give some additional (technical) considerations on the magnetic field evolution.

\section{Braking index measurements}

Braking index is a standard characteristic of radio pulsars rotational evolution (see, e.g., \citealt{1988MNRAS.234P..57B} and references therein). It is defined as:

\begin{equation}
n = \frac{\Omega \ddot \Omega}{\dot \Omega^2}=2 - \frac{P \ddot P}{\dot P^2}.
\label{e:brk_ind}
\end{equation}
Here $P$ and $\Omega$ are the spin period and frequency, respectively. 

Measuring braking indices is a complicated task, as it is necessary to obtain a robust value of the second derivative which is not just small, but also can be affected by microglitches and other types of timing noise. 
Recently, \cite{2019MNRAS.489.3810P, listBrakingInd} presented new measurements of braking indices for 19 young radio pulsars. 
The original sample of young objects with decade-long timing measurements used by \cite{2019MNRAS.489.3810P} included 85 sources.
 For eleven pulsars the obtained values are in the range $\sim 10$~--~$100$, two indices are negative, and two are above $100$. Only four values are compatible with the expected braking index $n=3$ predicted by the magneto-dipole formula for constant field and obliquity angle. 
For most of the 85 pulsars in the sample only upper limits  on the modulus of braking indices (in the range $n\lesssim 10$~--~$2000$) are obtained.

In \cite{2020MNRAS.tmp.2868I} we analysed several scenarios to explain these atypical values of the braking index. The main one is related to magnetic field decay. In the next section we briefly summarize basics of the field decay in NSs.

\section{Magnetic field decay}

There are many uncertainties in our understanding of magnetic field evolution (e.g., \citealt{2019LRCA....5....3P} and references therein). 
In the framework we use, the magnetic field decay in isolated radio pulsars proceeds at three distinctive stages (e.g., \citealt{IgoshevPopov2015} and references therein) driven by: (1) scattering of electrons off crystal phonons, which we refer to as ``phonon resistivity''; (2) Hall evolution; (3) resistivity due to crystal impurities in the crust. Correspondingly, one can introduce three timescales: $\tau_\mathrm{ph}, \tau_\mathrm{Hall}$, and $\tau_\mathrm{Q}$.

Timescale of the field decay due to phonon resistivity is related to NS thermal evolution. In the  so-called minimal cooling scenario (see e.g. \citealt{2004ApJS..155..623P})  the mean timescale seems to be around 400~kyr \citep{IgoshevPopov2015} (for this estimate, as well as for all our calculations, we use cooling curves obtained  by \citealt{Shternin2011_cooling_curves}). 

Hall evolution is not a dissipative process, it is related to changes in the magnetic field topology (see, e.g., \citealt{2018A&G....59e5.37G}).
The Hall time scale depends on the magnetic field of a NS as  \citep{cumming2004}:
\begin{equation}
\tau_\mathrm{Hall} = \frac{4\pi n_\mathrm{e} e L^2}{c B_\mathrm{p}}, \end{equation}
where  $c$ is the speed of light, $e$ is the elementary charge,  $B_\mathrm{p}$ is the poloidal dipolar magnetic field at the pole, $n_\mathrm{e}$ is the electron number density, and $L$ is length scale related to electric currents structure in the crust. The Hall evolution is dominant in magnetars due to their high magnetic fields.
 For a normal pulsar the Hall time scale is typically about a few Myrs.
 At $t\sim$~few~$\times \tau_\mathrm{Hall}$ the Hall cascade can saturate and thus, the configuration known as the Hall attractor is reached \citep{2014PhRvL.112q1101G}.
 
The timescale of decay due to crustal impurities seems to be the longest in normal pulsars. It exceeds at least 8~Myr and is compatible with the value even larger than 20~Myr \citep{Igoshev2019}.  
We want to underline that the evolution of the whole population of normal radio pulsars on timescale $\sim$ few tens of Myrs might proceed without continuous field decay which is characterized by a constant timescale as small as $\sim$ few $\times10^5$~yrs.
 This probably agrees with weak  dissipation due to impurities, since it operates along the whole life duration of an NS. Therefore,  crystal impurities cannot contribute significantly to field decay in all young pulsars, i.e., they cannot be responsible for the episode of decay discussed by \cite{igoshev2014, IgoshevPopov2015}. 
 
 Dissipation due to phonons and  impurities can be joined together to define the Ohmic time scale:

\begin{equation}
    \tau_\mathrm{Ohm}^{-1}=\tau_\mathrm{ph}^{-1} + \tau_\mathrm{Q}^{-1}.
    \label{e:tau_ohm}
\end{equation}
 Due to the field reconfiguration in the Hall cascade the length scale $L$ is decreased. Thus, the rate of dissipation can be significantly enhanced, as $\tau_\mathrm{Ohm}\propto L^2$ \citep{cumming2004}.

Magnetic field decay affects the first and the second period derivatives and therefore, affects both spin-down age and braking index and makes them correlated with each other. 

In the following subsection we describe some technical details of our approach to model the driving equation of the magnetic field evolution.

\subsection{Differential equation for magnetic field decay and its solution}

Calculations of poloidal dipolar magnetic field evolution in our models are based on the following equation \citep{aguilera2008}:

\begin{equation}
\frac{dB}{dt} = - \frac{B}{\tau_\mathrm{ohm}} - \frac{1}{B_0} \frac{B^2}{\tau_\mathrm{Hall}} .
\label{e:diff}
\end{equation}
Here we omit the lower index in notation of the magnetic field $B$. In this subsections we present a brief discussion of possible approaches to utilize this equation. 

If we assume that both $\tau_\mathrm{Ohm}$ and $\tau_\mathrm{Hall}$ are fixed (this can be valid, for example, for a short period of time), and $\tau_\mathrm{Hall}$ refers to the Hall timescale for the initial magnetic field, then 
the differential equation can be simplified by introducing auxiliary variables:
\begin{equation}
\alpha = \frac{1}{\tau_\mathrm{ohm}}    
\end{equation}
and
\begin{equation}
\beta = \frac{1}{B_0\tau_\mathrm{Hall}}   . 
\end{equation}
With this variables we obtain:
\begin{equation}
B'  + \alpha B = - \beta B^2 .
\end{equation}
First, we divide both sides to $B^2$:
\begin{equation}
B'B^{-2}  + \alpha B^{-1} = - \beta  
\end{equation}
This is the Bernoulli differential equation;
It can be solved using a substitution and integration multiplier.
We make a substitution: $w = 1/B$ and $w'=-B' / B^2$:
\begin{equation}
-w'  + \alpha w = - \beta  
\label{e:w}
\end{equation}
and use integration multiplier. In our case it is:
\begin{equation}
M(t) = \exp\left[-\int \alpha dt\right] =    \exp\left[- \alpha t\right].  
\end{equation}
We multiply both sides of eq.~(\ref{e:w}) to this multiplier and $(-1)$:
\begin{equation}
w'\exp\left[- \alpha t\right]  - \alpha w \exp\left[- \alpha t\right] = \beta \exp\left[- \alpha t\right] .
\label{e:e2}
\end{equation}
At this point we notice that:
\begin{equation}
(w\exp\left[- \alpha t\right])' = w'\exp\left[- \alpha t\right]  - \alpha w .
\end{equation}
Therefore we can easy integrate both sides of eq.~(\ref{e:e2}):
\begin{equation}
w \exp\left[- \alpha t\right] = - \frac{\beta}{\alpha} \exp\left[- \alpha t\right] + C     .
\end{equation}

After some transformations we return to the original variables:
\begin{equation}
B(t) = \frac{\exp\left[-t/\tau_\mathrm{Ohm}\right] }{C - \frac{\tau_\mathrm{Ohm}}{B_0\tau_\mathrm{Hall}} \exp\left[-t/\tau_\mathrm{Ohm}\right]}   .
\end{equation}
To obtain the value of $C$ we use the initial condition:
\begin{equation}
B(0) = B_0    ,
\end{equation}
which results into 
\begin{equation}
C = \frac{1}{B_0}\left(1 + \frac{\tau_\mathrm{Ohm}}{\tau_\mathrm{Hall}}\right).
\end{equation}
And we derive therefore eq. (17) from Aguilera et al. (2008) article:
\begin{equation}
B(t) = B_0\frac{\exp\left[-t/\tau_\mathrm{Ohm}\right] }{1 + \frac{\tau_\mathrm{Ohm}}{\tau_\mathrm{Hall}} - \frac{\tau_\mathrm{Ohm}}{\tau_\mathrm{Hall}} \exp\left[-t/\tau_\mathrm{Ohm}\right]}   .
\label{e:bt}
\end{equation}
If $\tau_\mathrm{ohm}$ evolves with the NS temperature, as in our model, we can solve the original eq. (\ref{e:diff}) numerically using e.g. Runge-Kutta integrator (as it was done in \citealt{2020MNRAS.tmp.2868I}), or use eq. (\ref{e:bt}) computing it at a fine grid and substituting value at the previous grid point instead of $B_0$, i.e.:
\begin{equation}
B_{i+1} = B_i \frac{\exp\left[-t/\tau_\mathrm{ohm}\right] }{1 + \frac{\tau_\mathrm{Ohm}}{\tau_\mathrm{Hall}} - \frac{\tau_\mathrm{Ohm}}{\tau_\mathrm{Hall}} \exp\left[-t/\tau_\mathrm{Ohm}\right]}    .  
\end{equation}
This  technique was used by us in \cite{igoshevpopov2018}.

\section{New results and comparison with our early studies}
\label{s:discussion}

In this section we, at first, briefly summarize results presented in \cite{2020MNRAS.tmp.2868I}, and then discuss how they are related with our previous modeling.

\begin{figure*}
    \centering
    \begin{minipage}{0.49\linewidth}
    \includegraphics[width=1.0\linewidth]{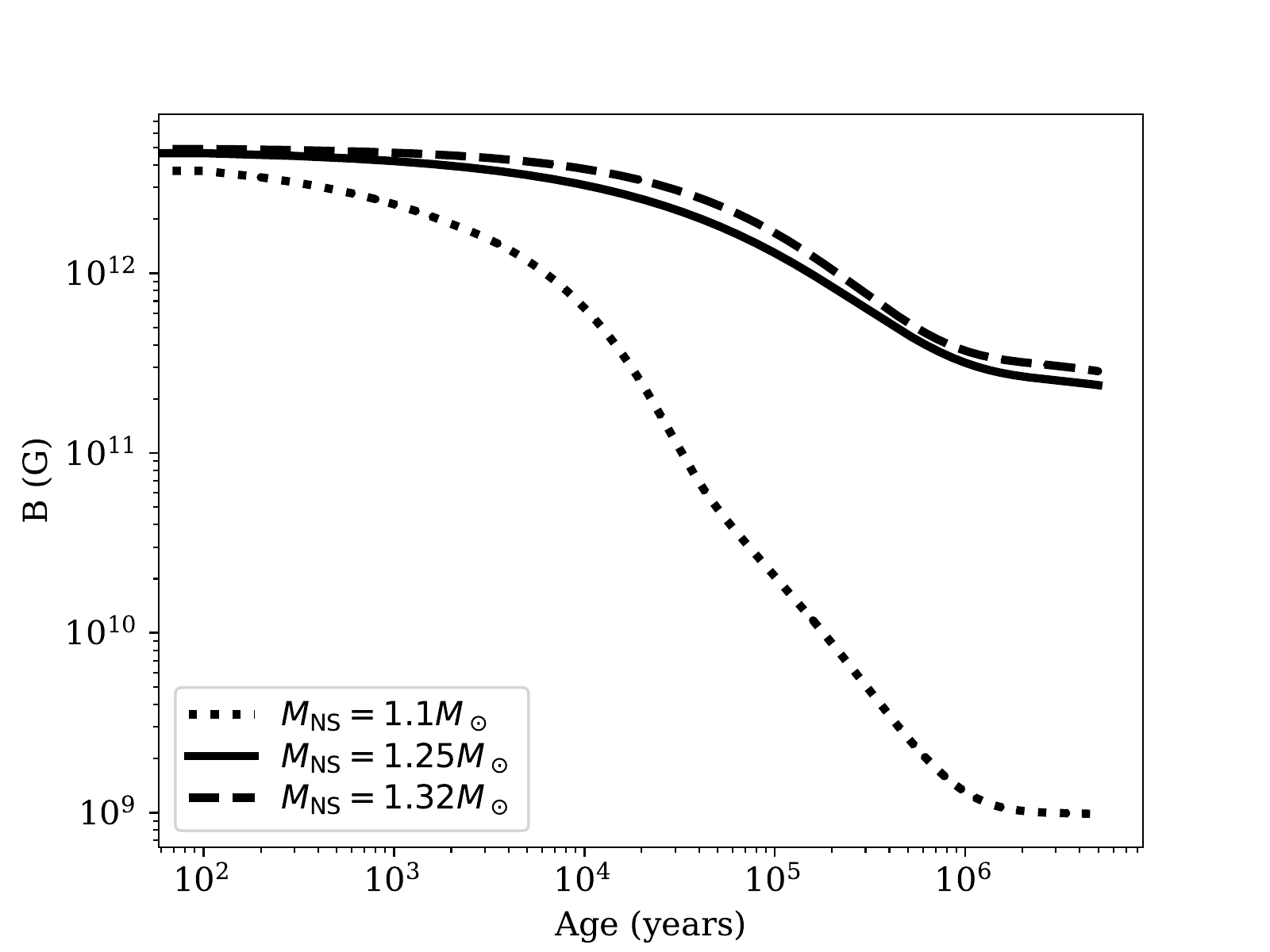}
    \end{minipage}
    \begin{minipage}{0.49\linewidth}
    \includegraphics[width=1.0\linewidth]{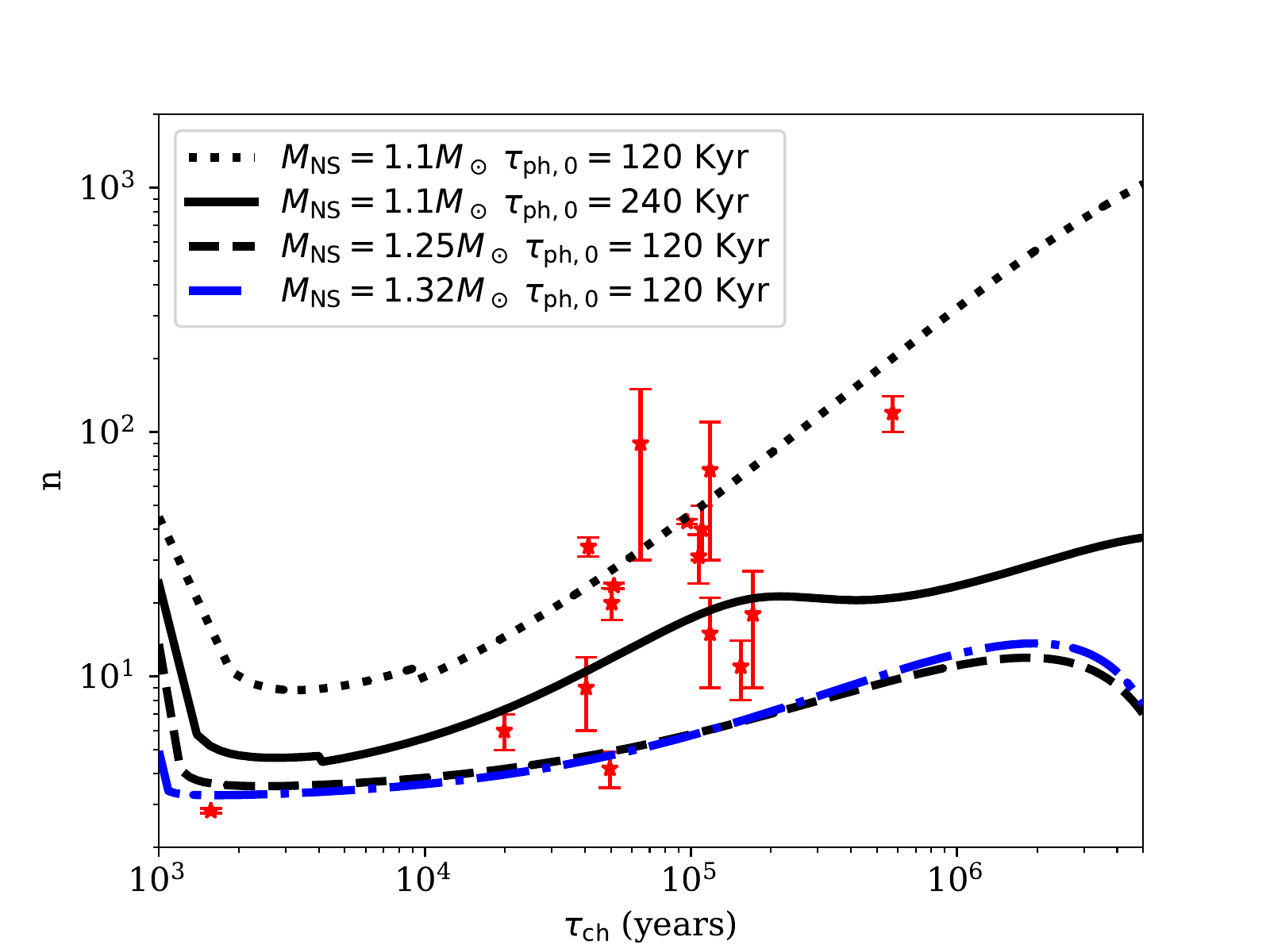}
    \end{minipage}    
    \caption{Left panel: magnetic field evolution for NSs of different masses, $\tau_\mathrm{ph,0}=1.2\times 10^5$~yrs and $\tau_Q = 200$~Myr. Right panel: dependence of braking index on spin-down age for different NS masses. Dots with error bars show measured values for 16 isolated radio pulsars from \protect\cite{listBrakingInd}. We fix $P_0 = 0.04$~s.}
    \label{f:dm}
\end{figure*}

The main results by \cite{2020MNRAS.tmp.2868I} are shown in Fig. \ref{f:dm} for a set of parameters slightly different from the one in the original paper. Increase of $\tau_\mathrm{ph, 0}$ by 20\% for the same set of masses resulted in just a slight decrease of predicted braking indices, which are still in the range occupied by the observed sample.

To parameterize dependence of $\tau_\mathrm{ph}$ on NS properties and internal  temperature we use the following equation: 
\begin{equation}
\tau_\mathrm{ph}
=\frac{\tau_\mathrm{ph, 0}}{T_8^2}    
\label{e:timescale}
\end{equation}
Here $\tau_\mathrm{ph, 0}$ in the first place depends on NS properties. 

In the right panel of Fig. \ref{f:dm} we see that atypical values of $n$ can be explained by low-mass NS. This is due to the fact that these compact objects cool slower, i.e. they can support intensive field decay due to phonon resistivity for a longer time.
Thus, one of the main conclusion made in \cite{2020MNRAS.tmp.2868I} is the following: braking indices in the range $\sim 10$~--~100 can be explained by field decay due to phonon resistivity in hot NSs. Later on, as these objects cool down, their braking indices might decrease and become close to classical value $\sim$ a few, and field decay proceeds due to crystal impurities in the crust. 

It is useful to examine if the fast magnetic field decay for the group of NSs proposed by \cite{2020MNRAS.tmp.2868I} is compatible with our earlier studies of young NS evolution
\citep{igoshev2014,IgoshevPopov2015}. In these earlier works we found that the dipolar magnetic field (sample averaged) decays approximately by the factor of two in the range of actual ages from $\sim$100~kyr up to $\sim400$~kyr, which roughly corresponds to exponential decay timescale of 400~kyr. This result was based on the cumulative distribution of spin-down ages for 516 isolated normal radio pulsars. 
Thus, a reasonable estimate for the fraction of pulsars with rapid decay is $\sim$20~per~cent, i.e., about the fraction of large braking indices seen in the sample by \cite{2019MNRAS.489.3810P}. However, note, that among the youngest sources in the sample the fraction becomes higher, so correspondingly we need larger fraction slowly cooling NSs to explain statistics of abnormal braking indices. 

In \cite{igoshev2014} we described the magnetic field decay as $B_\mathrm{p}(t) = B_0 \times f(t)$ where $B_\mathrm{p}(t)$ is time evolution of the poloidal dipolar magnetic field component, $B_0$ --- its initial strength, and $f(t)$ is the sample averaged magnetic field decay law. If the sample contains multiple groups of pulsars with individual decay laws $f_i(t)$ and fraction of each group is $w_i$, the sample average decay law is:

\begin{equation}
f(t) = \sum_{i=1}^N w_i f_i(t).    
\end{equation}

\begin{figure}
    \centering
    \includegraphics[width=1.0\linewidth]{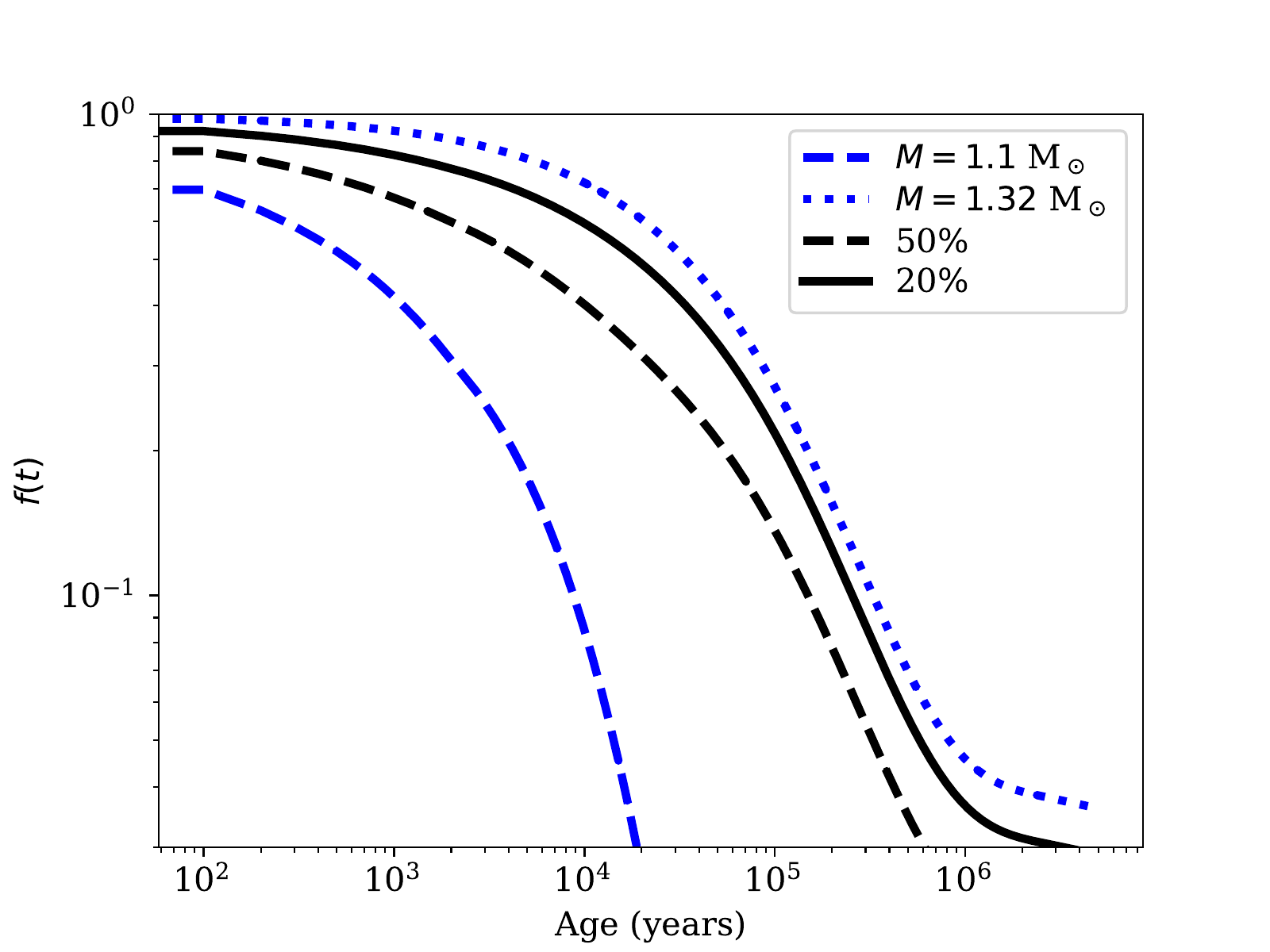}
    \caption{The dipolar poloidal magnetic field decay for individual NSs and sample average magnetic field decay law.  
    }
    \label{f:mixed}
\end{figure}

In Figure~\ref{f:mixed} we show sample averaged decay law for 20~per~cent and 50~per~cent fractions of low-mass NSs ($M = 1.1$~$M_\odot$) in comparison with field evolution for two individual masses (again cooling curves by \cite{Shternin2011_cooling_curves} are used; note that results on field evolution depend on the thermal history and other parameters of NSs, so here our calculations might be taken mostly as an illustration). 

As it is seen from the Figure, the sample averaged decay closely follows the curve typical for NSs with moderate magnetic field decay (here we take the mass $1.32\, M_\odot$ as an example).\footnote{Decay might be much
slower for very massive NSs with $M \gtrsim 1.8$~$M_\odot$ --- for the chosen equation of state, --- with efficient direct URCA cooling, in these objects the phonon resistivity becomes unimportant very rapidly.} 

The sample averaged magnetic field reaches the value $f=0.32$ in 100~kyr and $f=0.14$ in 400~kyr. It means a decay by factor $\sim2.3$ proceeds during the time period from 100 kyr to 400 kyr --- compatible with our previous findings. 

Here we can make two conclusions: \\
(1) our method presented in \cite{igoshev2014} is just weakly sensitive to presence of a small NS group  with fast magnetic field decay; \\
(2) fast magnetic field decay for the case of a $\sim$20~per~cent hot NS-group is compatible with results obtained by \cite{igoshev2014} for a large set of young radio pulsars selected only by their characteristic ages.  For slightly larger fractions of hot NSs results are still marginaly compatible.

Note, that in this analysis we do not try to obtain the best fit for $\tau_\mathrm{ph,0}$ which could explain all available data. This is a task for future studies using detailed population modelling (with realistic NS mass distribution, different cooling histories for all different masses, etc.). Instead, we simply demonstrate that fast decay of magnetic field in $\gtrsim$20~per~cent of NSs explains new measurements of braking indices and simultaneously is in reasonable correspondence with earlier studies of field decay in young radio pulsars.

Recently, \cite{gusakov2020} suggested that an episode of magnetic field decay at ages $\sim 10^5$~yrs can be explained by  field evolution in a NS core. This is an interesting and intriguing possibility. Only recently realistic models of field behaviour in the core started to appear. We hope to explore applications of such models to the population of known radio pulsars in near future.



\section{Conclusions}

Modeling of NS magnetic field evolution and comparison of theoretical results with observational data continue to points towards existence of an episode of a brief field decay  at ages $\sim$~few hundred kyrs with a characteristic time scale also about few~$\times100$~kyrs. 
Recently, we demonstrated \citep{2020MNRAS.tmp.2868I} that such an episode can explain an abundance of positive, anomalous braking indices of some radio pulsars, and that such field decay can be attributed just to a fraction of young NSs (about a few tens of per cent) which stay hot for necessarily long time. This result is compatible with our early findings \citep{igoshev2014, IgoshevPopov2015}. In this short note we summarized these results, presented examples for slightly different sets of parameters in comparison with \cite{2020MNRAS.tmp.2868I}, and discussed compatibility of the new and previous calculations.

\section*{Acknowledgements}

SP thanks Peter Shternin for comments on cooling curves.
AI thanks support from the Science and Technology Facilities Council for research grant ST/S000275/1.  Work of SP is supported by the Ministry of science and higher education of Russian Federation under the contract 075-15-2020-778 in the framework of the Large scientific projects program within the national project "Science" and by the program of leading scientific schools of the Lomonosov Moscow State university.

\bibliography{popov_igoshev_AN}

\end{document}